\documentstyle[12pt,epsf]{article}

\newcommand{\e}{{\it e}}
\newcommand{\m}{\overline{m}}
\newcommand{\beq}{ \begin{eqnarray} }
\newcommand{\eeq}{ \end{eqnarray} }
\newcommand{\beqstar}{ \begin{eqnarray*} }
\newcommand{\eeqstar}{ \end{eqnarray*} }
\newcommand{\gsim}{ \mathop{}_{\textstyle \sim}^{\textstyle >} }
\newcommand{\lsim}{ \mathop{}_{\textstyle \sim}^{\textstyle <} }

\begin{document}
\baselineskip 0.7cm

\begin{titlepage}

\begin{center}

\hfill LBL-38653, UT-751\\
\hfill TU-502, TUM-HEP-246/96\\
\hfill SFB-375/99\\
\hfill hep-ph/9605296\\
\hfill \today

  {\large \bf Exact Event Rates of Lepton Flavor Violating Processes
    \\ in Supersymmetric $SU(5)$ Model } \vskip 0.5in {\large
    J.~Hisano$^{(a)}$\footnote
{Fellows
of the Japan Society for the Promotion of Science.}\footnote
{The address after the September 1996 is,
School of Physics and Astronomy, 
University of Minnesota, 
Minneapolis, MN 55455, USA.},
 T.~Moroi$^{(b)}$, K.~Tobe$^{(a,c)*}$ and
    M.~Yamaguchi$^{(d)}$\footnote
{On leave of absence from Department of
    Physics, Tohoku University, Sendai 980-77, Japan.}
    } 
\vskip 0.4cm {\it (a) Department of Physics, University of
    Tokyo, Tokyo 113, Japan}
\\
{\it (b) Theoretical Physics Group, Lawrence Berkeley Laboratory, \\ 
  University of California, Berkeley, CA 94720, U.S.A.}
\\
{\it (c) Department of Physics, Tohoku University, Sendai 980-77,
  Japan}
\\
{\it (d) Institute f\"{u}r Theoretische Physik \\ Physik Department,
  Technische Universit\"{a}t M\"{u}nchen \\ D-85747 Garching, Germany}
\vskip 0.5in

\abstract {Event rates of various lepton flavor violating processes in the
  minimal supersymmetric $SU(5)$ model are calculated, using exact formulas
  which include Yukawa vertices of lepton-slepton-Higgsino.  We find subtlety
  in evaluating event rates due to partial cancellation between diagrams.  This
  cancellation typically reduces the event rates significantly, and the size of
  the reduction strongly depends on superparticle mass spectrum. }
\end{center}
\end{titlepage}

\setcounter{footnote}{0}

Lepton Flavor Violation (LFV) is an indication of physics beyond the standard
model.  Supersymmetric (SUSY) extension of the standard model, which has been
strongly motivated as a solution of the gauge hierarchy problem, generally
possesses LFV as generation mixing of slepton masses~\cite{EN}.  In particular,
interaction of superheavy particles may induce LFV in the slepton masses
through renormalization effects, giving non-negligible event rates for LFV
processes~\cite{HKR}.  Concrete realization of the mechanism has been
considered in supersymmetric grand unified theories (GUTs) such as $SU(5)$
\cite{HKR,BH,BHS}, $SO(10)$ \cite{BHS,CRS,ACH} and flipped $SU(5)$ \cite{GM},
as well as in models with right-handed neutrinos \cite{BM,HMTYY,HMTY}.

A remarkable observation has been made in Refs.~\cite{BH,BHS}, which
has pointed out that large top-quark Yukawa coupling yields large
rates of LFV processes in SUSY GUTs only one or two orders of
magnitudes below present experimental bounds.  The large top Yukawa
coupling gives a sizable flavor violation in slepton masses squared
through the renormalization effects above the GUT scale.  Importantly
\cite{BH,BHS}, the amount of the off-diagonal elements in the slepton
mass matrix has mild dependence on the value of the top Yukawa
coupling even when it closes to its quasi-infrared fixed point.

However, this stable prediction of the LFV source in the slepton masses will
not necessarily lead to a stable prediction of event rates.  The purpose of
this paper is to give exact event rates of $\mu \rightarrow \e \gamma$, $\mu
\rightarrow \e\e\e$, $\mu$-$\e$ conversion in nuclei, and $\tau
\rightarrow \mu \gamma$ in the case of the minimal SUSY $SU(5)$ model, and show
that this is the case.  When calculating these event rates, we include Yukawa
type vertices of lepton-slepton-Higgsino~\cite{HMTY}, which was not taken into
account in Ref.~\cite{BHS}.  As was already suggested in Ref.~\cite{HMTY}, new
diagrams coming from the Yukawa vertices interfere destructively to the other
dominant diagrams, leading to reduction of the event rates.  We find that this
partial cancellation is subtle and generally has complicated dependence on
superparticle mass spectrum.  It takes place significantly under the universal
scalar mass hypothesis because contribution from each diagram is comparable,
and event rates we obtain are typically much smaller than the ones previously
analyzed \cite{BH,BHS}.  Moreover, we will show that,  given a LFV scalar mass
matrix, variation of other parameters within a reasonable range will change the
event rates by one order of magnitude or more.  Without detailed knowledge on
superparticle mass spectrum, it would be difficult to predict these LFV event 
rates.

First of all, let us introduce the model we consider.
In the minimal SUSY $SU(5)$ model both quarks and leptons are embedded
in the {\bf 5}$^*$ and {\bf 10} representations.
The left-handed ($SU(2)_L$ doublet) leptons are in {\bf 5}$^*$'s, while the
right-handed ($SU(2)_L$ singlet) ones are in {\bf 10}'s.  The 
Yukawa interaction 
in this model is described by a superpotential
\begin{eqnarray}
W &=& \frac14 f_{\rm u}^{ij} \Psi_i \Psi_j H 
+ \sqrt{2}f_{\rm d}^{ij} \Psi_i \Phi_j \overline{H}
\label{Yukawa}
\end{eqnarray}
with $f_{\rm u}^{ij}$ and $f_{\rm d}^{ij}$ being Yukawa couplings.  Here
supermultiplets $\Psi_i$ and $\Phi_i$ correspond to the {\bf 10} and {\bf
  5}$^*$ representations, respectively, and indices $i$ and $j$ ($i,j=1, 2, 3$) stand
for a generation.  $H$ ($\overline{H}$) is a Higgs supermultiplet in {\bf 5}
({\bf 5}$^*$) representation.  In the decomposition of the standard model,
$f_{\rm u}$ becomes Yukawa coupling matrix of the up-type quark sector, whereas
$f_{\rm d}$ involves those of the down-type quark sector and of the lepton
sector.  The Yukawa couplings in Eq.~(\ref{Yukawa}) break lepton-flavor number
conservation as well as its quark counter part: one can diagonalize one of the
Yukawa couplings matrices, $f_{\rm d}$ for example, with respect to generation
indices.  When it is done, the other Yukawa couplings $f_{\rm u}$ among the
{\bf 10}'s cannot be diagonalized in general and thus yield flavor violation.
In the model we are considering, we assume that this is the only source of LFV.

Above the GUT scale $M_G \approx 2 \times 10^{16}$ GeV, 
soft SUSY breaking terms for  scalar fields are given by
\begin{eqnarray}
-{\cal L}_{\rm SUSY breaking} 
&=&
  (m_{10}^2)_{\; i}^j \psi^{\dagger i}\psi_{j}
+ (m_{5}^2)_i^{\; j} \phi^{\dagger i}\phi_{j}
+ m_{h}^2 h^{\dagger}h
+ m_{\overline{h}}^2 \overline{h}^{\dagger}\overline{h}
\nonumber\\
&&+ \Bigl(
\frac14 A_{\rm u}^{ij} \psi_i \psi_j h + \sqrt{2} A_{\rm d}^{ij} 
\psi_i \phi_j \overline{h}
+ h.c. \Bigr),
\label{soft-terms}
\end{eqnarray}
where we have used lowercases $\psi_i$, $\phi_i$, $h$, and $\overline h$ to
denote scalar components of the supermultiplets $\Psi_i$, $\Phi_i$, $H$, and
$\overline H$, respectively.  Soft masses for right-handed sleptons are
identified with $m_{10}^2$ and those for left-handed sleptons are $m_5^2$.  At
the (reduced) Planck scale, $M \approx 2 \times 10^{18}$ GeV, it is assumed
that the matrices $m_{10}^2$ and $m_{5}^2$ are diagonal and $A_{\rm u}$ and
$A_{\rm d}$ are proportional to the Yukawa couplings $f_{\rm u}$ and $f_{\rm
d}$, respectively, in order to avoid too large flavor changing neutral
currents. When the energy scale goes down below the Planck scale, they will
acquire flavor-violating components through renormalization effects which
involve the flavor-violating Yukawa couplings. The magnitude of the LFV in the
soft masses can be evaluated by using renormalization group equations above the
GUT scale.  At one-loop level, some of the equations are written as
\begin{eqnarray}
 16 \pi^2 Q \frac{d}{d Q} (m_{10}^2)_{\; i}^j 
&=& 
-  \frac{144}{5} g_5^2 M_5^2 \delta_{\; i}^j
+  \left\{   2 f_{\rm d} f_{\rm d}^{\dagger} 
           + 3 f_{\rm u} f_{\rm u}^{\dagger}, m_{10}^2\right\}_{\; i}^j
\nonumber\\
&&
+  4\left[   (f_{\rm d} f_{\rm d}^\dagger)_{\; i}^j m_{\overline{h}}^2
           + (f_{\rm d} m_5^2 f_{\rm d}^\dagger 
        + A_{\rm d} A_{\rm d}^\dagger)_{\; i}^j \right]
\nonumber\\
&&
+  6\left[   (f_{\rm u} f_{\rm u}^\dagger)_{\; i}^j m_{h}^2
           + (f_{\rm u} m_{10}^2 f_{\rm u}^\dagger 
         + A_{\rm u} A_{\rm u}^\dagger)_{\; i}^j \right],   
\\
  16 \pi^2 Q \frac{d}{d Q} (m_{5}^2)_i^{\; j} 
&=&  
-  \frac{96}{5} g_5^2 M_5^2 \delta_i^{\; j}
+  4 \left\{f_{\rm d}^{\dagger} f_{\rm d}, m_{5}^2\right\}_i^{\; j}
\nonumber\\
&&
+  8 \left[   (f_{\rm d}^\dagger f_{\rm d})_i^{\; j} m_{\overline{h}}^2
           + (f_{\rm d}^\dagger m_{10}^2 f_{\rm d} 
         + A_{\rm d}^\dagger A_{\rm d})_i^{\; j} \right],
\\
 16 \pi^2 Q \frac{d}{d Q} A_{\rm u}^{ij}
&=&
\left[ -\frac{96}{5} g_5^2 + 3 {\rm tr}(f_{\rm u} f_{\rm u}^{\dagger})
\right] A_{\rm u}^{ij} 
+2 \left[
-\frac{96}{5} g_5^2 M_5 + 3 {\rm tr}(A_{\rm u} f_{\rm u}^{\dagger})
\right] f_{\rm u}^{ij}
\nonumber \\
&&+ (2 f_{\rm d} f_{\rm d}^{\dagger} 
  + 3 f_{\rm u} f_{\rm u}^{\dagger})^i_{\; k} A_{\rm u}^{kj}
+(2f_{\rm d} f_{\rm d}^{\dagger} + 3 f_{\rm u} f_{\rm u}^{\dagger})^j_{\; k}
A_{\rm u}^{ik}
\nonumber \\
&&+ 2(2 A_{\rm d} f_{\rm d}^{\dagger} 
+ 3 A_{\rm u} f_{\rm u}^{\dagger})^i_{\; k} f_{\rm u}^{kj}
+2(2A_{\rm d} f_{\rm d}^{\dagger} + 3 A_{\rm u} f_{\rm u}^{\dagger})^j_{\; k}
f_{\rm u}^{ik},
\\
16 \pi^2 Q \frac{d}{d Q} A_{\rm d}^{ij}
&=&
\left[ -\frac{84}{5} g_5^2 + 4 {\rm tr}(f_{\rm d} f_{\rm d}^{\dagger})
\right] A_{\rm d}^{ij} 
+2 \left[
-\frac{84}{5} g_5^2 M_5 + 4 {\rm tr}(A_{\rm d} f_{\rm d}^{\dagger})
\right] f_{\rm d}^{ij}
\nonumber \\
&&+ (10 f_{\rm d} f_{\rm d}^{\dagger} 
  + 3 f_{\rm u} f_{\rm u}^{\dagger})^i_{\; k} A_{\rm d}^{kj}
+(8 A_{\rm d} f_{\rm d}^{\dagger} + 6 A_{\rm u} f_{\rm u}^{\dagger})^i_{\; k}
f_{\rm d}^{kj}.
\end{eqnarray}
In these equations, $g_5$ is the gauge coupling constant of the unified gauge
group, $M_5$ is the corresponding gaugino mass and $Q$ stands for the
renormalization point (energy scale).
 
Solutions of the one-loop renormalization group equations above the
GUT scale were discussed already \cite{BHS}.  Following
Ref.~\cite{BHS}, let us take the basis where the Yukawa matrix $f_{\rm
u}$ is diagonalized. Dropping the Yukawa couplings and $A$-parameters
except for $f_{\rm u}^{33}(\equiv f_t)$ and $A_{\rm u}^{33}$ ({\it i.e.} those for
top quark) in the renormalization group equations, one can solve them
analytically.\footnote 
{The renormalization effect by the bottom Yukawa coupling
decreases flavor-diagonal, third generation, masses of the left- and
right-handed sleptons, but does not significantly induce LFV masses. This could
be easily understood in the basis that $f_{\rm d}$ is diagonal.  Though we will
include the non-zero bottom Yukawa coupling in our numerical analysis, for the
moment we ignore it for simplicity.}
It turns out that the universality for the flavor
indices no longer holds in $m_{10}^2$, $A_{\rm u}$, and $A_{\rm d}$,
while it holds in $m_5^2$. In particular, the matrix $m_{10}^2$ at
the GUT scale is generally written as
\begin{eqnarray}
  \left(
  \begin{array}{ccc}
  m^2 & 0        & 0          \\
    0      & m^2 & 0          \\
    0      &   0      &  m^2-I 
  \end{array}
  \right),
\end{eqnarray}
where $m^2$ is a generation blind contribution and $I$ stands for a
contribution from the top Yukawa coupling.  Below the GUT scale, it is more
convenient to use the basis in which the Yukawa matrix for leptons are
diagonalized. Such a basis can be obtained by using the Kobayashi-Maskawa (KM)
matrix at the GUT scale. In this basis, soft masses for the right-handed
sleptons exhibit LFV and their amounts are proportional to a bilinear of elements
of the KM matrix ($V$):
\begin{equation}
      (m_{\tilde l_R}^2)_{\;i}^{j}=m^2 \delta_{\;i}^{j}- V_{\;i}^{3*} V_{3}^{\;j} I 
~~~~(Q\simeq M_G).
\end{equation}

If universality of soft masses is imposed at the Planck scale, the quantity $I$
is compactly expressed in the minimal SUSY $SU(5)$ model as~\cite{BHS}
\begin{eqnarray}
     I&= & \rho_G [m_0^2+\frac{1}{3} (1 -\rho_G) A_0^2 
               +0.198 (1 - \rho_G) A_0 M_{5G} 
\nonumber \\
     & & +(0.224 -0.029 \rho_G) M_{5G}^{\; 2}].
\label{expr-I}
\end{eqnarray}
$M_{5G}$ is a gaugino mass at the GUT
scale. Precise meaning of an universal scalar mass $m_0$ and an universal
$A$ parameter $A_0$ at the Planck scale will be given below. 
In the above equation, $\rho_G$ is defined as
\begin{equation}
      \rho_G=\left( \frac{f_t(M_G)}{f_t^{\rm max}(M_G)}
             \right)^2
\end{equation}
where $f_t^{\rm max}(M_G)$, which equals 1.56 in the minimal model, is a
maximal value of the top Yukawa coupling at the GUT scale under the assumption
that the theory is perturbative below the Planck scale.   Conversely if one
starts with a large value of the top Yukawa coupling at the Planck scale, it
converges to $f_t^{\rm max} (M_G)$ at the GUT scale and hence it exhibits a
property as a (quasi) infrared fixed point. 
Thus
$\rho_G \leq 1$ and equality holds if the Yukawa coupling is on the fixed
point.  The structure of Eq.~(\ref{expr-I}) is simple. It is proportional to
$\rho_G$ as a whole, and the quantity in the parenthesis is of the order of the
SUSY breaking scale.  Furthermore as the top Yukawa coupling gets close to the
maximal value, $I$ converges almost to a value $m_0^2$.  This simplicity is an
important finding of Refs.~\cite{BH,BHS}.

For the reader's convenience, we present in Fig.~1 dependence of the LFV source
on the top quark mass.  The horizontal line is  the top pole mass
($m_t^{\rm pole}$) divided by  $\sin\beta$, which is defined as 
$\tan\beta=\langle h_2 \rangle/\langle h_1\rangle$ with $h_1$ ($h_2$) 
being the Higgs boson that give masses to  down-type quarks
  and charged leptons (up-type quarks). The vertical
line represents  $(m^2_{\tilde l_R})_e^{\mu}/m^2_{\tilde e_{R}}$ where 
$(m^2_{\tilde l_R})_e^{\mu}$ is a $(\mu, e)$ component of the 
right-handed slepton masses squared, and $m_{\tilde e_R}$ is a 
physical right-handed selectron mass. These are
evaluated at an electroweak scale, {\it i.e.},
at the $Z$-boson mass. Here we choose $M_{5G}$ and $m_0$  such that a Bino 
mass $M_1$
is 50GeV and $m_{{\tilde e}_R}=300$ GeV, and $A_0=0$. The real line, 
dashed line, and dash-dotted line correspond to the case for 
$\tan \beta=30,10,3$, respectively.

Complete formulas of the event rates for the LFV processes were given
in Ref.~\cite{HMTY} where we took a mass-eigenstate basis for the
sleptons.  Instead of repeating them, we would like to discuss the
qualitative behavior of the LFV rates using an approximate
mass-insertion formula. In the approximation, the mass-eigenstate
basis for the leptons is taken, and LFV masses squared as well as
left-right mixing slepton masses are treated as perturbation.  (This
approach gives us a good approximation when the generation mixing in
the chirality conserving scalar masses dominates over that in the $A$
parameters. This is checked numerically in the model we are
considering.)  In this discussion, we concentrate on $\mu \rightarrow
e \gamma$.  The process is described by effective
electromagnetic-dipole type matrix element:
\begin{eqnarray}
  T=e \epsilon^{\alpha *}\bar{u}_\e m_{\mu} i \sigma_{\alpha \beta}
  q^\beta (A_2^L P_L + A_2^R P_R) u_{\mu},
\label{Penguin}
\end{eqnarray}
where  $P_{R/L}=(1\pm\gamma_5)/2$, $m_\mu$ is a muon mass, $e$ the 
electric charge,
$q$ a photon momentum, and $\epsilon^{\alpha}$ is the photon
polarization vector. From this equation, the event rate of $\mu\rightarrow 
e \gamma$ is given as
\begin{eqnarray}
\Gamma(\mu \rightarrow e~\gamma)
= \frac{e^2}{16 \pi} m_{\mu}^5 (|A_2^L|^2+|A_2^R|^2).
\end{eqnarray}
In the  model we are discussing,
$A_2^R$ is almost zero since left-handed sleptons 
have only tiny LFV soft-breaking masses with the non-vanishing bottom Yukawa
coupling. The diagrams contributing to this
matrix element have to have chirality flip of lepton and, to be proportional
to $SU(2)_L\times U(1)_Y$ breaking vacuum expectation values. Such
diagrams in this model are shown in Figs.~2. Fig.~2(a), in which the chirality
flips on an external line, gives the following contribution to
$A_2^L$,\footnote
{In our case, diagrams with two mass insertions like 
$(m_{\tilde{l}_R}^2)^{\mu}_{\tau} (m_{\tilde{l}_R}^2)^{\tau}_{e}$ could give a
non-negligible contribution.  However, we do not incorporate such  higher terms
since they do not change our argument drastically. Especially they do not
change the sign of each contribution discussed here and subsequently.}
\begin{eqnarray}
  A_2^{L}|_{\rm a} &=& -\frac16 \frac{\alpha_Y}{4 \pi}
  \frac{1}{\m_{\tilde{l}_R}^2} \frac{(m_{\tilde{l}_R}^2)_{\rm
      e}^{\mu}}{\m_{\tilde{l}_R}^2},
\end{eqnarray}
where, $\m_{\tilde{l}_R}$ is an averaged right-handed charged slepton
mass, $(m_{\tilde{l}_R}^2)_{\rm e}^{\mu}$ a ($\mu,\e$) component of the 
right-handed charged slepton mass squared matrix, 
and $\alpha_Y\equiv g_Y^2/4
\pi$ with $g_Y$ being the $U(1)_Y$ gauge coupling constant.
To make our points clearer, here we adopt an expansion with respect to
$M_1^2/\m_{\tilde{l}_R}^2$, though we will show the exact results in
the following figures. This expansion is motivated by the fact that the
right-handed slepton masses become larger than the Bino mass through
the renormalization effects, unless sfermion masses are negative at
the Planck scale. (For example, $\m_{\tilde{e}_R} \gsim 2 M_1$ in the
case of the minimal SUSY $SU(5)$ model.)

In Figs.~2(b), (c) and (d), the chirality flips on an internal line. 
Fig.~2(b) gives the following contribution,
\begin{eqnarray}
  A_2^{L}|_{\rm b} &=& \frac12 \frac{\alpha_Y}{4 \pi}
  \frac{1}{\m_{\tilde{l}_R}^2} \frac{(m_{\tilde{l}_R}^2)_{\rm
      e}^{\mu}}{\m_{\tilde{l}_R}^2} \frac{M_1(A_\mu/f_\mu + \mu \tan\beta
    )}{\m_{\tilde{l}_L}^2}.
\label{a2fig2}
\end{eqnarray}
In this equation $\m_{\tilde{l}_L}$ is an averaged left-handed charged
slepton mass, $A_\mu$ a SUSY-breaking trilinear coupling proportional
to muon Yukawa coupling constant $f_\mu$.  It is notable that
$A_2^{L}|_{\rm b}$ grows as $\tan\beta$, and can dominate over
$A_2^{L}|_{\rm a}$ if $3M_1\mu\tan\beta /\m_{\tilde{l}_L}^2\gsim 1$.
Remember that matrix element Eq.~(\ref{Penguin}) is proportional to 
the Yukawa coupling of the
lepton, and to one of  $SU(2)_L\times U(1)_Y$ breaking vacuum expectation
values, $\langle h_1\rangle$ or $\langle h_2\rangle$. Some diagrams 
hitting the vacuum expectation value of $h_2$ result in the enhancement
if $\tan\beta$ is large~\cite{HMTYY,HMTY}.

Once the Yukawa couplings of lepton-slepton-Higgsino are 
correctly included,
Figs.~2 (c) and (d) also contribute to $\mu \rightarrow \e \gamma$ as
\begin{eqnarray}
A_2^{L}|_{\rm c} &=& - \frac{\alpha_Y}{4 \pi}   
  \frac{1}{\m_{\tilde{l}_R}^2}   
  \frac{(m_{\tilde{l}_R}^2)_{\rm e}^{\mu}}{\m_{\tilde{l}_R}^2}
  \frac{M_1 \mu \tan\beta}{\m_{\tilde{l}_R}^2}  
f_1( \frac{\mu^2}{\m_{\tilde{l}_R}^2}),
\\
A_2^{L}|_{\rm d} &=& \frac{\alpha_Y}{4 \pi}   
  \frac{1}{\m_{\tilde{l}_R}^2}   
  \frac{(m_{\tilde{l}_R}^2)_{\rm e}^{\mu}}{\m_{\tilde{l}_R}^2}
   f_2( \frac{\mu^2}{\m_{\tilde{l}_R}^2}),
\label{a2fig3}
\end{eqnarray}
with 
\begin{eqnarray}
f_1(x) &=& - \frac{8 - 11 x + 4 x^2 - x^3 + 2(2 + x) \log x}{2(1 - x)^4},\\
f_2(x) &=& \frac{1 + 4 x - 5 x^2 + 2x(2 + x)\log x}{2( 1 - x)^4}.
\end{eqnarray}
Functions $f_1(x)$ and $f_2(x)$ are monotonously decreasing functions
of $x$ and positive-definite. As $x$ goes infinity, both functions
vanish. Thus $A_2^{L}|_{\rm c}$ and $A_2^{L}|_{\rm d}$ are not
negligible unless $|\mu|$ is much larger than $\m_{\tilde{l}_R}$. 
As a result, 
$A_2^L|_{\rm b}$ and $A_2^L|_{\rm c}$ are
important when $M_1 \mu \tan \beta$ is large, otherwise $A_2^L|_{\rm
a}$ and $A_2^L|_{\rm d}$ will dominate. An important point is that
the relative signs between $A_2^{L}|_{\rm b}$ and $A_2^{L}|_{\rm
c}$, and that between $A_2^{L}|_{\rm a}$ and $A_2^{L}|_{\rm d}$ are both
opposite.  This fact implies possible cancellation between the
diagrams.  We will soon show that this indeed occurs.

We are now at a position to present our result of numerical
computation using the formulas in Ref.~\cite{HMTY}. 
First, we consider the case where all  scalars have a common soft  mass
at the reduced Planck scale ({\it i.e.}, a universal scalar mass).
The model is characterized by the following parameters at the
scale $M$: a universal scalar mass $m_0$, a $SU(5)$ gaugino mass
$M_{5}$, a universal trilinear scalar coupling $A_0$, a mixing
mass parameter of the two Higgs scalars $B$, and a supersymmetric
Higgsino mass parameter $\mu$. Then, the boundary conditions for the
parameters in Eq.~(\ref{soft-terms}) are given by
$(m_{10}^2)^{\; i}_j=(m_5^2)^i_{\; j} =m_0^2 \delta^i_j$, $m_h^2=m_{\overline
h}^2=m_0^2$, $A_{\rm u}^{ij}=f_{\rm u}^{ij}A_0$, and $A_{\rm d}^{ij}=f_{\rm d}^{ij}A_0$
at the Planck scale.
Two of the five parameters (usually, $B$ and $|\mu|$)
are fixed to obtain the desired values of the
$Z$-boson mass and $\tan\beta$.  

In this analysis, we fix the top Yukawa coupling at the Planck scale to be
$f_t(M)=2.4$ which corresponds to, at the GUT scale, $f_t(M_G)=1.4$. We
choose this value partly for comparison with Ref.~\cite{BHS}. It gives the top
mass $m_t \simeq$ 170 (180, 190) GeV for $\tan\beta =2$ (3, 10).\footnote
{The Yukawa coupling constant we adopt may result in a top quark mass which is 
fairly large. However, we shall note here that the event rate becomes smaller
for a smaller value of $f_t$. In that case, experimental verification of LFV 
becomes severer than the case discussed in this paper.}
Also we choose $A_0=0$.  Dependence of the size of the generation mixing 
in the slepton masses squared on these parameters was discussed earlier.

Branching ratio for $\mu \rightarrow e \gamma$ is shown in Fig.~3.
The horizontal line in the figure is taken to be the physical right-handed
selectron mass $m_{\tilde e_R}$, and $M_{5}$ is fixed such that the
Bino mass $M_1$ is 50 GeV. In our numerical analysis, we imposed
theoretical and experimental constraints, {\it i.e.}, negative
searches for the SUSY particles at LEP1 and Tevatron~\cite{PDG}, a
muon anomalous magnetic dipole moment~\cite{muon} ($-1.34\times
10^{-8}\leq \frac{1}{2}(g-2)\leq 2.34\times 10^{-8}$), and tree-level 
stability of the scalar potential along $h_1=h_2$ direction. 
We present 
results in the case of $\tan \beta=3$, 10, 30 for each sign of
$\mu$.

As can be seen from this figure, there exists a region where the
cancellation between the diagrams significantly reduces the event rate
of $\mu \rightarrow e \gamma$. With the universal scalar mass
hypothesis, an absolute value of Higgsino mass $\mu$ determined by the
$SU(2)_L\times U(1)_Y$ symmetry breaking condition increases with
$m_{\tilde e_R}$ in a range 180GeV $\lsim |\mu| \lsim $ 200GeV for
$\tan\beta$=10 or 30, $M_1=50$ GeV and $m_{\tilde e_R} \lsim 300$ GeV
(200GeV $\lsim |\mu| \lsim $ 240GeV for $\tan\beta$=3).  Since the
mass range for $\mu$ is comparable with other superparticle masses, we
cannot ignore the diagrams which have the Yukawa type vertices.  In a
small $m_{\tilde e_R}$ region $A_2^{L}|_{\rm b}$ tends to dominate
over other contributions, and in a large $m_{\tilde e_R}$ region
$A_2^{L}|_{\rm c}$ does, especially if $\tan\beta$ is large. Therefore,
in a middle, there is steep cancellation. If the Yukawa coupling of
Higgsino were ignored, the branching ratio would exceed $10^{-11}$ in
the case, for example, when $\tan\beta=30$ and $m_{\tilde
e_R}<200$GeV. Under the universal scalar mass hypothesis, however, the
cancellation between diagrams reduces the branching ratio to $\sim
10^{-12}$.\footnote
{The event rate of $\mu\rightarrow e \gamma$ does not drop off as quickly as 
the fourth power of the right-handed selectron mass above the
cancellation point, but has rather a flat
behavior. Since the renormalization effects by the top and bottom Yukawa
couplings reduce the right-handed stau mass significantly, the loop diagram
like Fig.2 (c), which is proportional to $(m_{\tilde{l}_R}^2)^{\mu}_{\tau}
(m_{\tilde{l}_R}^2)^{\tau}_{e}$, dominates the other diagrams completely when
the right-handed selectron mass is larger than the bino and Higgsino mass.
While this fact does not change the qualitative behavior that there is a
cancellation between diagrams as we mentioned in the previous footnote, the
quantitative behavior in that region changes even with right-handed
selectron mass larger \cite{HT}. The other LFV events also have such smooth
behavior as that of $\mu\rightarrow e \gamma$.}
Also, in the region of valleys $A_2^{L}|_{\rm a}$ and 
$A_2^{L}|_{\rm d}$ contribute to the amplitude with the same order as 
the sum of $A_2^{L}|_{\rm b}$ and $A_2^{L}|_{\rm c}$, and
this leads to the difference of positions of valleys of each line.
Moreover, if the top
Yukawa coupling is smaller, the radiative symmetry breaking condition
requires smaller values of $|\mu|$, and the point of cancellation
tends to shift to a region with smaller right-handed selectron mass. Thus, the
situation is more complicated than that previously considered in
Ref.~\cite{BHS}.

Next, we show branching ratios for $\mu \rightarrow \e\e\e$, $\mu\mbox{-}\e$
conversion in $^{48}_{22}$Ti, and $\tau \rightarrow \mu \gamma$ in Figs.4-6.
Configurations of these figures are the same as Fig. 3. These events are reduced
by about one order of magnitude compared to the case where the Yukawa vertices
are not taken into account, and these processes also suffer from
the partial cancellation.

The behavior of the $\mu \rightarrow eee$ is similar to that of $\mu
\rightarrow e \gamma$ up to normalization.  This is because the penguin type
contribution from Eq.~(\ref{Penguin}) is enhanced by phase space integral and
hence it dominates over other contributions\cite{HMTY}. Steepness of the
valleys is somewhat less in the case of $\mu \rightarrow eee$, because
diagrams other than the penguin diagram contribute.  On the other hand, the
enhancement by the phase space integral does not occur for the $\mu$-$e$
conversion and thus behavior of this event rate is not similar to that of $\mu
\rightarrow e \gamma$. 
Then, the cancellation occurs at right-handed selectron mass different 
from the case of $\mu \rightarrow e \gamma$ and this suggests search of 
$\mu$-$e$ conversion would have a complementary role with that of $\mu
\rightarrow e \gamma$. 
Also $\tau \rightarrow \mu \gamma$ behaves like $\mu \rightarrow e \gamma$ 
does and they are related with each other through the KM matrix.

So far, we have assumed the universal scalar mass at the Planck scale.
It is well-known, however, the degeneracy of all soft scalar masses is
not needed to avoid disastrously large flavor changing neutral
currents.  In the context of $SU(5)$, the mass degeneracy of the
sfermions with the same quantum numbers ({\bf 5$^*$} and {\bf 10}) is
sufficient.  Thus we shall discuss the $\mu \rightarrow \e \gamma$
rate as an example in a non-universal case.

To study the case with non-universality, we show how the results
depend on the mass spectrum of the superparticles with a fixed value of
the right-handed mass matrix
\begin{eqnarray}
m^2_{{\tilde l}_R}=m^2_{\tilde e_R}
\left (
\begin{array}{ccc}
  1.0        & -9.8 \times 10^{-5}     & -3.2 \times 10^{-3} \\
 -9.8 \times 10^{-5}      & 1.0        & -2.3 \times 10^{-2} \\
 -3.2 \times 10^{-3}   & -2.3 \times 10^{-2}     & 0.25 \\
\end{array}
\right).
\end{eqnarray}
(These values are realized for the universal mass case with $m_{\tilde
e_R}=300$ GeV, $M_1=50$ GeV, $A_0=0$, $\tan\beta=3$ and $f_t(M)=2.4$.)
In Figs.~7 and 8, branching ratio for the process $\mu \rightarrow \e \gamma$ 
is shown as a function of the physical left-handed selectron mass
$m_{{\tilde e}_L}$ and $\mu$-parameter.
Here we take $M_1=50$ GeV, $m_{{\tilde e}_R}=300$ GeV, and
$\tan \beta=3$ (Fig.~7) and 30 (Fig.~8). We assume that all left-handed
charged slepton have a common mass $m_{{\tilde e}_L}$ and  $A(m_Z)=0$
for simplicity. The shaded regions are excluded by the present experiments.

In general, the amplitude for the $\mu \rightarrow e \gamma$ is a complicated
function of parameters at low energy scale. In addition to the right-handed
slepton masses squared, it depends on the following parameters at the
electroweak scale: $m_{\tilde l_L}$, $A$, $\mu$, $\tan \beta$, $M_1$ (and
$M_2$).  However, the qualitative behavior can be easily understood by using
the mass insertion formula discussed above. 
For example, the left-handed slepton
mass becomes smaller or Higgsino mass $\mu$ larger, $A_2^{L}|_{\rm b}$ in
Eq.~(\ref{a2fig2}) becomes larger than the others.  When the left-handed
slepton mass becomes larger, $A_2^{L}|_{\rm a}$  and $A_2^{L}|_{\rm d}$ 
( $A_2^{L}|_{\rm c}$) becomes dominant for $\tan\beta=3$ ($\tan\beta=30$). 
Thus,
the branching ratio becomes larger because the cancellation does not occur
significantly.

We should emphasize that our low-energy approach given here should apply to a
certain class of the unified models where the right-handed slepton masses are
the only source of flavor violation.  The effect of this interference can be
important when one computes the event rate in a given model and we expect that
the rates suffer from this complicated cancellation in a large class of models.
On the other hand, there is not such significant cancellation in models which
left-handed sleptons have LFV masses, {\it e.g.}, $SO(10)$ SUSY GUT 
\cite{BHS,CRS,ACH} and the minimal SUSY standard model with 
right-handed neutrinos \cite{BM,HMTYY,HMTY} since 
the LFV process is usually dominated by only one diagram.

\section*{Acknowledgment}
Two of the authors (J.H. and K.T.) would like to thank Y.~Okada and
T.~Yanagida for useful discussion. The work of T.M. was supported by
the Director, Office of Energy Research, Office of High Energy and
Nuclear Physics, Division of High Energy Physics of the U.S.
Department of Energy under Contract DE-AC03-76SF00098, and the work of
M.Y. was supported by the Sonderforschungsbereich 375-95: ``Research
in Particle-Astrophysics" of the Deutsche Forschungsgemeinschaft and
the ECC under contracts No.  SC1-CT91-0729 and No. SC1-CT92-0789.

\newpage
%
%
\newcommand{\Journal}[4]{{\sl #1} {\bf #2} {(#3)} {#4}}
\newcommand{\APJ}{Ap. J.}
\newcommand{\CJP}{Can. J. Phys.}
\newcommand{\NC}{Nuovo Cimento}
\newcommand{\NP}{Nucl. Phys.}
\newcommand{\PL}{Phys. Lett.}
\newcommand{\PR}{Phys. Rev.}
\newcommand{\PRep}{Phys. Rep.}
\newcommand{\PRL}{Phys. Rev. Lett.}
\newcommand{\PTP}{Prog. Theor. Phys.}
\newcommand{\SJNP}{Sov. J. Nucl. Phys.}
\newcommand{\ZP}{Z. Phys.}

\newpage
%
%
%
%
\begin{figure}[p]
\epsfxsize=15cm
\caption
{Top quark mass dependence of flavor-violating right handed slepton mass
$({m_{{\tilde l}_R}^2})^\mu_e$ at the electroweak scale.
Real line, dashed line, and dash-dotted line correspond to the case for
$\tan \beta=30,10,3$. Here we take $M_1=50$ GeV and $m_{{\tilde e}_R}=300$ GeV,
$A_0=0$. We represent the points corresponding to $f_t(M)=0.5,1.0,
2.0,2.4,5.0$, which is the top Yukawa coupling constant at the reduced 
Planck scale.}
\end{figure}
%
%
\begin{figure}
\epsfxsize=15cm
\caption
{Feynman diagrams contributing to the process $\mu\rightarrow e \gamma$
in the minimal SUSY $SU(5)$ GUT.
The symbols $\tilde{\mu}_{L}$, $\tilde{\mu}_{R}$, $\tilde {e}_{R}$, 
$\tilde{B}^0$, and $\tilde{H}^0$ represent left-handed smuon, 
right-handed smuon, right-handed selectron, Bino, and neutral Higgsino 
respectively.  The blobs in the slepton line indicate the insertion of the
flavor-violating right-handed slepton mass $({m_{{\tilde l}_R}^2})^\mu_e$
and left-right mixing mass which is proportional to the vacuum expectation
value $\langle h_1\rangle$ or $\langle h_2\rangle$. 
The blobs in the Bino-Higgsino line 
represent the insertion of gaugino-Higgsino mass mixing, that is,
$\mu$ denotes Higgsino $(\tilde{H}_1-\tilde{H}_2)$ mass mixing, 
$\langle h_{1,2}\rangle$ 
Higgsino-Bino $(\tilde{H}^0-\tilde{B}^0)$ mass mixing,
$M_1$ Bino mass. And blob in the external muon line represent the chirality 
flip of the external muon $\mu$.}
\end{figure}
%
%
%
%
\begin{figure}[p]
\epsfxsize=15cm
\caption
{Branching ratios for the process $\mu \rightarrow e \gamma$ as a function
of the physical right-handed selectron mass,
$m_{{\tilde e}_R}$. Real line, dotted line, dash-dotted line correspond
to the case for the $\tan \beta=30,10,3$ respectively. Dashed line represents
the present experimental upper bound for this process. Here we take 
$M_1=50$ GeV, $f_t(M)=2.4$, and (a) $\mu > 0$ (b) $\mu < 0$.}
\end{figure}
%
%
%
%
%
\begin{figure}[p]
\epsfxsize=15cm
\caption
{Branching ratios for the process $\mu \rightarrow eee$.
Here we take the same parameters as Fig.3.}
\end{figure}
%
%
%
%
\begin{figure}[p]
\epsfxsize=15cm
\caption
{The $\mu$-$e$ conversion rates in nuclei $^{48}_{22}$Ti.
Here we take the same parameters as Fig.3.}
\end{figure}
%
%
%
%
\begin{figure}[p]
\epsfxsize=15cm
\caption
{Branching ratios for the process $\tau \rightarrow \mu \gamma$.
Here we take the same parameters as Fig.3.}
\end{figure}
%
%
%
%
\begin{figure}[p]
\epsfxsize=15cm
\caption
{Branching ratios for the process $\mu \rightarrow \e \gamma$ as a
  function of the physical left-handed selectron mass $m_{{\tilde
      e}_L}$ and $\mu$-parameter for the case of non-universal scalar
  mass. Here we take $M_1=50$ GeV, $m_{{\tilde e}_R}=300$ GeV,
  $f_t(M)=2.4$, $\tan \beta=3$, $A(m_Z)=0$ and the typical value of
  right-handed slepton mass matrix as mentioned in text. For simplify
  we assume that all left-handed charged slepton have a common mass
  $m_{{\tilde e}_L}$. The shaded regions are excluded by the present 
  experiments.} 
\end{figure} 
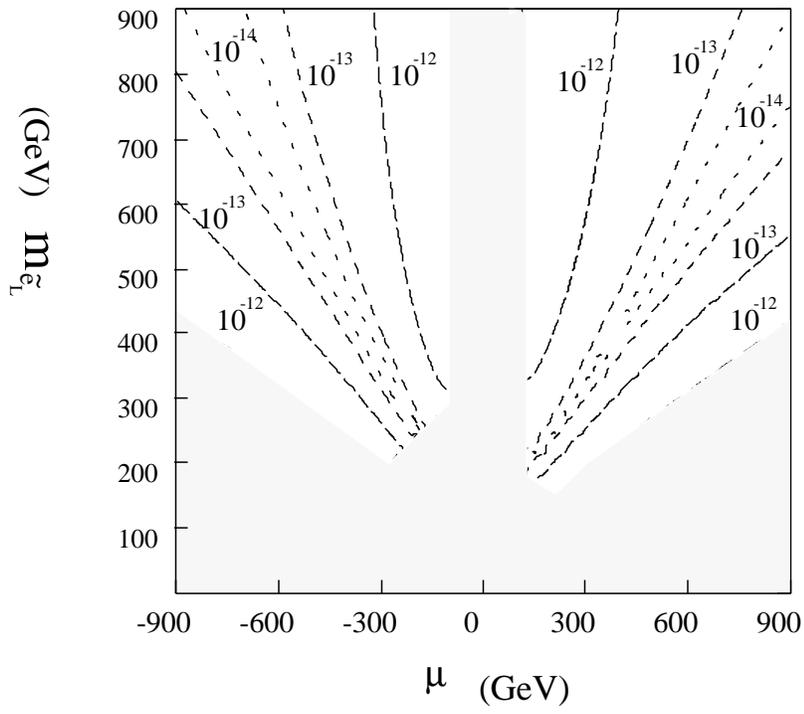
\begin{figure}
[p] 
\epsfxsize=15cm 
\centerline{} 
\caption 
{Same as  Fig.7 except for $\tan \beta=30$.} 
\end{figure} 
\end{document}